# Spin Accumulation in Nondegenerate and Heavily Doped p-Type Germanium

Satoshi Iba[*], Hidekazu Saito, Aurelie Spiesser, Suguru Watanabe[1], Ron Jansen,

Shinji Yuasa, and Koji Ando

National Institute of Advanced Industrial Science and Technology (AIST),

Spintronics Research Center, Tsukuba, Ibaraki 305-8568, Japan

[1]Tsukuba University, Tsukuba, Ibaraki 305-8571, Japan

**Abstract**

Spin accumulation induced in p-type germanium from Fe/MgO tunnel contacts is studied as a function of hole concentration $p$ ($10^{16}$ - $10^{19}$ $cm^{-3}$). For all $p$, the contacts are free of rectification and Schottky barrier, guaranteeing spin injection into the Ge and preventing spin accumulation enhancement by two-step tunneling via interface states. The observed spin accumulation is smallest for nondegenerate doping ($p \sim 10^{16}$ $cm^{-3}$) and increases for heavily doped Ge. This trend is opposite to what is expected from spin injection and diffusion theory. For heavily doped Ge, the observed spin accumulation is orders of magnitude larger than predicted.

[*] E-mail address: s.iba@aist.go.jp

The electrical creation and detection of spin-polarized carriers in a semiconductor (SC) is an important topic in the research field of spintronics. To date, the creation and detection of spin by tunneling from a ferromagnetic contact have been achieved in GaAs,[1-3] Si[4-9] and Ge[10-14]. However, what controls the magnitude of the spin accumulation induced in the SC is not understood. Tran *et al.*[2] noted that the spin signal in $Co/Al_2O_3/GaAs$ devices is orders of magnitude larger than what is expected from the standard theory for spin injection and spin diffusion in the SC[15-17]. They proposed two-step tunneling via localized states near the SC interface and described a model that predicts that spin accumulation in such localized states can be greatly enhanced[2]. However, the validity of the model for their system was not tested or confirmed. Dash *et al.*[5] observed spin accumulation in $Ni_{80}Fe_{20}/Al_2O_3/Si$ devices and found spin signals several orders of magnitude larger than expected. Importantly, Dash *et al.* performed an explicit test[5] that showed that enhancement by two-step tunneling is not the origin of the large signals observed at room temperature, implying that there must be another cause of the discrepancy between experiment and theory. Spin accumulation much larger than predicted has now been observed in GaAs, Si, Ge, in n-type and p-type materials, and with amorphous ($Al_2O_3$, $SiO_2$) and crystalline (MgO) tunnel barriers[2,5,7,9,10,12,13]. In all these cases, the SC is heavily doped (above the metal-insulator transition). Interestingly, Ando *et al.*[8] reported that the spin accumulation in nondegenerately doped Si with CoFe/Si Schottky contacts is not enhanced and can be well explained by the standard theory.

Here, we report spin accumulation versus doping density in a single material system, namely, p-type Ge with crystalline Fe/MgO tunnel contacts, varying the doping density from nondegenerate ($< 10^{16}$ cm$^{-3}$) to heavy doping ($> 10^{18}$ cm$^{-3}$) above the metal-insulator transition. The scaling of the magnitude of the spin accumulation with doping concentration is compared to spin injection and diffusion theory. We find that the observed spin accumulation is smallest for nondegenerate doping ($p \sim 10^{16}$ cm$^{-3}$) and increases for heavily doped Ge. This trend is opposite to what is expected from theory, and for heavily doped Ge, the observed spin accumulation is orders of magnitude larger than predicted.

The p-type Ge was chosen because, as shown below, it allows the fabrication of tunnel contacts that are free from carrier depletion and Schottky barrier formation in the SC.[10] This guarantees that injected spins accumulate in the Ge and prevents the build-up of an enhanced spin accumulation in interface states, which are always present, because enhancement can only[2] exist if the localized interface states are decoupled from the bulk SC bands by a (Schottky) tunnel barrier with a sufficiently large resistance $r_b$. The magnitude of the spin signal has an upper limit that is set by the value of $r_b$, and in the absence of a Schottky barrier, the spin accumulation in the interface states is equal to that in the bulk bands of the SC.[2] The absence of a Schottky barrier also avoids nonlinear, rectifying, and thermally assisted transport, for which data cannot be compared, as recently done for n-type Ge,[18] with spin injection theory[15-17] derived for tunneling in linear response.

Films were grown by molecular beam epitaxy on p-type Ge(001) substrates with different hole concentrations as listed in Table I. Tunnel contacts consisting of Au (20 nm)/Fe (10 nm)/MgO (2.0-2.5 nm) were deposited at room temperature, following the previously described procedure.[10] Since there is no Schottky barrier, the use of a graded doping profile at the surface was not needed (not even for the nondegenerately doped substrate), thus avoiding complications due to the formation of a potential well in the SC.[19] Junctions with an active tunnel area ($A$) of $100 \times 200$ $\mu m^2$ were prepared as previously described.[10] The spin accumulation was probed at a temperature ($T$) of 40 K using Hanle and inverted Hanle measurements in a three-terminal configuration.[1,5,10,20]

The hole concentration $p$ and resistivity $\rho_{Ge}$ of the substrates are given in Fig. 1(a) and Table I. At $T$ = 300 K, the substrates have $p = 8 \times 10^{18}$, $4 \times 10^{17}$, and $1 \times 10^{16}$ $cm^{-3}$ and will be referred to using labels HI, ME, and LO, respectively, denoting high, medium, and low doping. Substrate LO exhibits a rapid decrease of $p$ below about 30 K, consistent with the nondegenerate character. Substrates HI and ME have a weak $T$ dependence down to 5 K, owing to conduction in a metallic impurity band at high acceptor density.[21,22)]

Current-voltage (*I-V*) characteristics of the Fe/MgO/Ge contacts are presented in Figs. 1(b) and 1(c). If a Schottky barrier would be present, one expects diode behavior, with a smaller current for negative bias voltage, particularly at low *T*. Moreover, rectification would be very pronounced for low doping concentration for which the depletion region would be wide. However, we observe similar *I-V* characteristics without any diode behavior for all the contacts, even at 40 K and for contacts prepared on the LO substrates. Also, the junction resistance *R* has a weak *T* dependence - for example, *R*(40K)/*R*(280K) is about 2.5 at 10 mV bias. All this demonstrates that the MgO/Ge contacts are free from a Schottky barrier. Note that this result is expected to depend on the fabrication procedure, and different conditions may lead to a more pronounced diode behavior. [23)]

Figures 2(a) and 2(b) present Hanle and inverted Hanle curves at 40 K for the samples on HI- and ME-doped substrates, obtained with the magnetic field applied perpendicular ($B_{\perp}$) and parallel ($B_{||}$) to the tunnel interface, respectively. Clear Lorentzian line shapes centered around zero field can be seen, proving that a spin accumulation has been induced in the Ge by tunnel injection of spins from the Fe contact. The magnitude of the spin signal ($\Delta V_{spin}$) corresponds to the sum of the voltage change in the Lorentzian part of the Hanle and inverted Hanle curves[10,20)]. The values are about 200 μV for sample HI and 330 μV for sample ME. As shown in Fig. 3(a), the spin signal for sample LO was overshadowed by a large positive background due to Lorentz magnetoresistance (LMR) in the Ge substrate, which was also observed in control devices using a nonmagnetic Au contact. This LMR is larger for the LO substrate owing to its larger mobility (LMR is quadratic in mobility), and because the larger substrate resistance is a larger fraction (10%) of the total three-terminal resistance. Nevertheless, clear deviations from a quadratic function are found in the low *B* range, suggesting that a spin accumulation is also induced in the LO substrate. To further visualize the spin signal, the voltage difference between the Hanle and inverted Hanle data [$V(B_{\perp})$ - $V(B_{||})$] is given in Fig. 3(b). A peak centered at zero with a Lorentzian line shape is observed, but only for the device with the Fe contact, indicating that a spin accumulation is created in the nondegenerate Ge. The corresponding $\Delta V_{\rm spin}$ is about 80 μV.

The observed $\Delta V_{\rm spin}$ is converted to a spin resistance-area-product [spin-*RA* ≡ ($\Delta V_{\rm spin}$ /J), where *J* is the current density *I/A*], and plotted in Fig. 4 (solid red circles). The spin-*RA* of the ME device (5.1 kΩμm$^2$) is close to the previously reported value (~ 6 kΩμm$^2$ at 40 K),[10] indicating good reproducibility of the spin-*RA* values. The experimental data are compared with the standard theory for spin injection and spin diffusion,[15-17] which predicts a spin-*RA* equal to $P^2 \cdot \rho_{\rm Ge} \cdot \lambda_{\rm sd}$, where *P* is the tunnel spin polarization (~0.6 in our Fe/MgO contacts[10]) and $\lambda_{\rm sd}$ is the spin-diffusion length. The predicted spin-*RA* versus *p* is also plotted in Fig. 4 (open blue circles), for different values of $\lambda_{\rm sd}$. For the nondegenerate (LO) device, the measured spin-*RA* matches the theoretical value if $\lambda_{\rm sd}$ = 2 μm. From the width of the Hanle curve of the LO sample [Fig. 3(a)] we extract a spin lifetime $\tau_s$ > 43 ps. This is a lower limit due to the presence of artificial broadening of the Hanle curve.[20] Using $\lambda_{\rm sd} = (D \cdot \tau_s)^{1/2}$ and a diffusion constant $D$ = 21 cm$^2$/s determined from the measured mobility at 40 K, we obtain a lower limit for $\lambda_{\rm sd}$ of 0.3 μm. Agreement requires $\tau_s$ = 1.9 ns, which is not inconsistent with $\tau_s$ > 43 ps. However, we note that the spin-*RA* obtained at 20 K [Fig. 3(b)] is comparable to that at 40 K, whereas the theory predicts that it should be a factor of 5 larger because $\rho_{\rm Ge}$ at 20 K is a factor of 5 larger, and *P* and $\lambda_{\rm sd}$ are not expected to be smaller at 20 K. This indicates that for the nondegenerate sample, the data is not in agreement with the theory.

For the heavily doped devices, the measured spin-*RA* is larger than that of the LO device. However, theory predicts the opposite trend, because the heavily doped samples have a much lower $\rho_{\rm Ge}$ (Table I) and $\lambda_{\rm sd}$ is expected to be shorter for higher *p*. For the heavily doped devices, a match of experiment and theory requires an unrealistically large value of $\lambda_{\rm sd}$ (610 and 43 μm for samples HI and ME, respectively). In other words, the measured spin-*RA* for the heavily doped Ge is about 2-3 orders of magnitude larger than the theory predicts.

As discussed, the absence of a Schottky barrier (negligible $r_b$) means that an enhancement of the spin accumulation by interface states cannot be present. Moreover, the increasing discrepancy between theory and experiment at higher *p* is inconsistent with such an enhancement. The $r_b$ of any residual Schottky barrier decreases rapidly with increasing *p* due to the shrinkage of the depletion region and enhancement by interface states, if present, would be less pronounced at a high doping density. The measured spin-*RA* thus represents the spin accumulation in the Ge bulk bands. For nondegenerate doping, the observed spin accumulation is closest to the prediction of the theory of spin injection and detection, but the variation with temperature is not according to the expectation. For heavily doped p-type Ge, the theory severely underestimates the magnitude of the induced spin accumulation. The origin of the discrepancy between the experiment and theory is not understood and further investigations are needed. The demonstration that the data and theory exhibit the opposite variations with *p* provides an important new piece of information.

**Acknowledgments**

We are grateful to S. Sharma, M. Yamamoto, and K. Moriyama (National Institute of Advanced Industrial Science and Technology) for help with sample fabrication and measurements. This work was supported by a Funding Program for Next Generation World-Leading Researchers (No. GR099). One of the authors (A. S.) acknowledges JSPS Postdoctoral Fellowship for Foreign Researchers.

**Figure captions**

FIG. 1 (a) Hole concentration versus temperature for p-type Ge substrates with heavy, medium, and low doping, labeled HE, ME, and LO, respectively. (b) Current-voltage characteristics at 40 K for Fe/MgO/Ge tunnel contacts on the three different substrates. (c) The same at 300 K. The bias voltage is defined as $V_{Ge} - V_{Fe}$, where $V_{Ge}$ and $V_{Fe}$ are the potentials of the Ge and Fe electrodes, respectively.

FIG. 2 Hanle and inverted Hanle curves at 40 K for Fe/MgO/Ge devices with (a) HI- and (b) ME-doped Ge substrates, as indicated. A constant current was applied such that holes tunnel from the Ge substrate into Fe, with the current adjusted to obtain the same bias voltage (200 mV) for both devices. Magnetic fields were applied perpendicular ($B_{\perp}$, red) and parallel ($B_{||}$, blue) to the film plane for Hanle and inverted Hanle curves, respectively.

FIG. 3 (a) Same type of data as in Fig. 2, but for devices with LO-doped Ge substrate with a Fe/MgO contact (solid circles), and Au/MgO contact (open circles), at 40 K. Solid lines are fits to a parabolic background. (b) Voltage difference $V(B_{\perp}) - V(B_{||})$ for the devices with LO doping as a function of magnetic field, obtained from the data of Fig. 3(a). For Fe contacts, data at 20 K is also shown.

FIG. 4 Spin-*RA* ($\Delta V_{spin}$ divided by current density) of the Fe/MgO/Ge devices measured at 40 K (solid red circles), as a function of the hole concentration of the Ge. Also shown is the value predicted by spin injection/diffusion theory[16-18)] (open blue circles) for different values of the spin-diffusion length $\lambda_{sd}$, as indicated.

Table I. Doping element, hole concentration *p*, and resistivity $\rho_{Ge}$ of the p-type Ge substrates, together with the spin-*RA* measured in Fe/MgO/Ge tunnel devices, all at 40 K.

| Sample | Dopant | *p* at 40 K ($cm^{-3}$) | $\rho_{Ge}$ at 40 K (mΩcm) | Spin-*RA* at 40 K (kΩμ$m^2$) |
|---|---|---|---|---|
| HI | Ga | $7.4 \times 10^{18}$ | 2.3 | 5.1 |
| ME | In | $4.4 \times 10^{17}$ | 33 | 5.1 |
| LO | In | $4.9 \times 10^{15}$ | 210 | 1.5 |

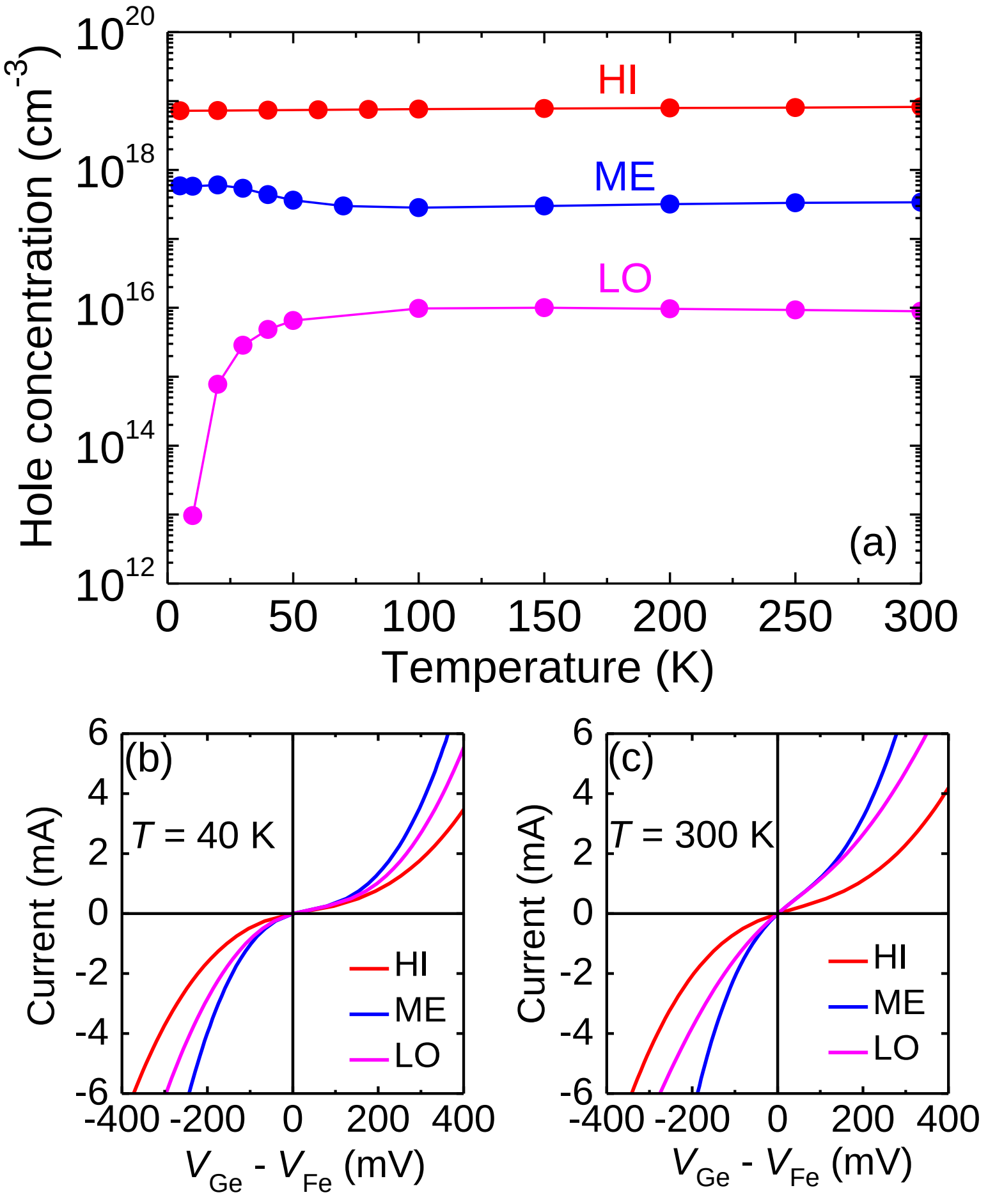

Figs.1 (a)-(c) Iba *et al.*

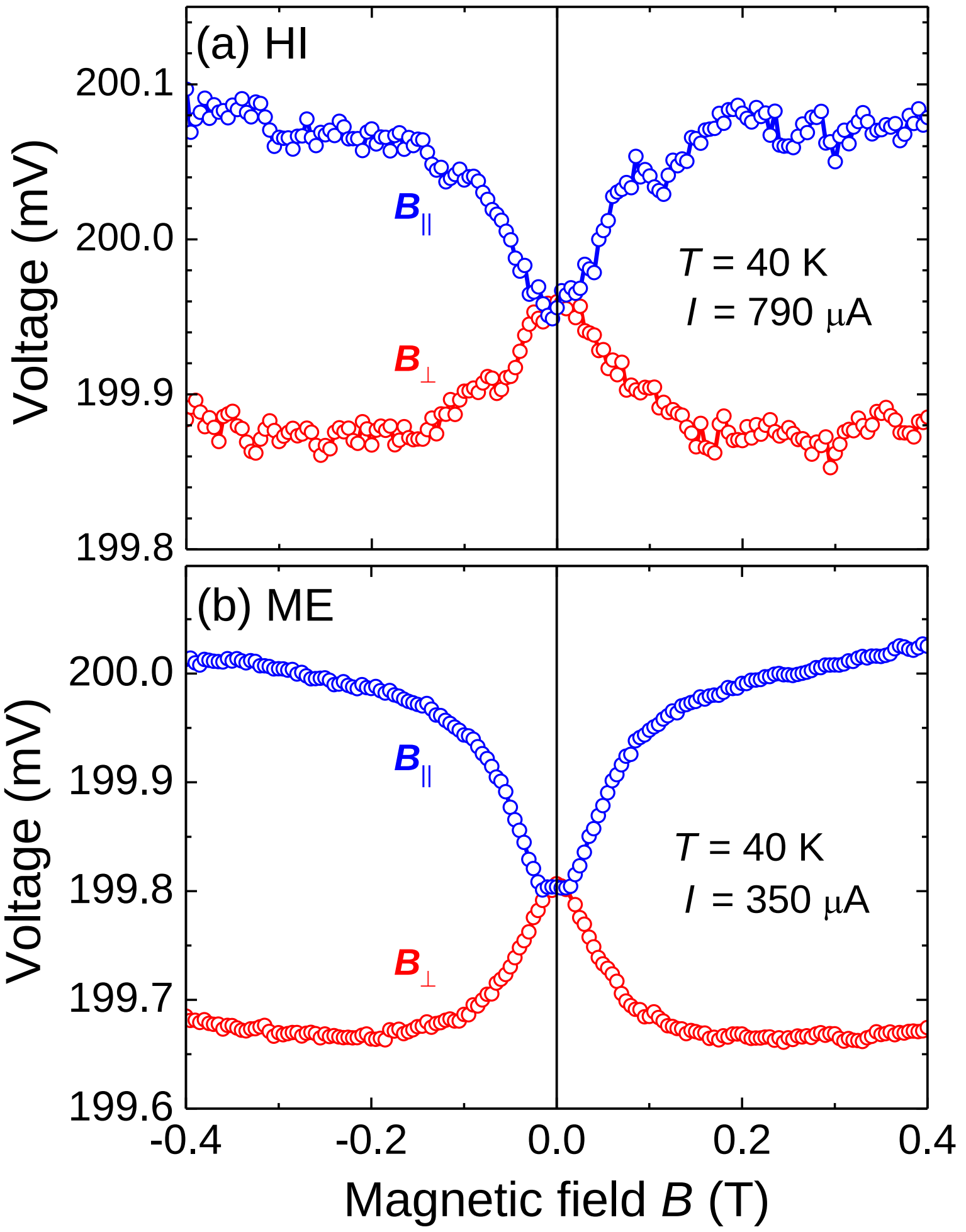

Figs.2 (a) and (b) Iba *et al*.

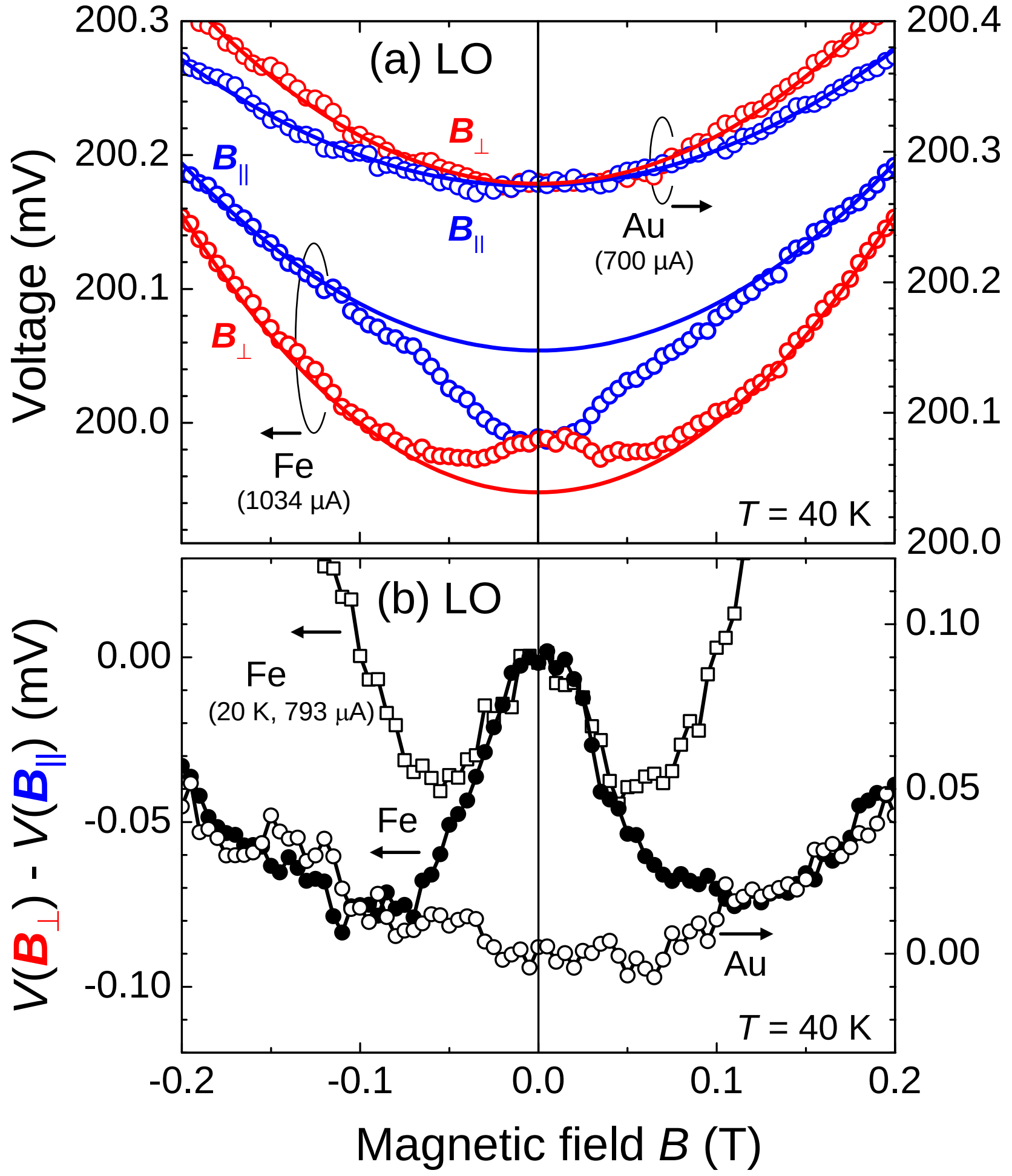

Figs.3 (a) and (b) Iba *et al*.

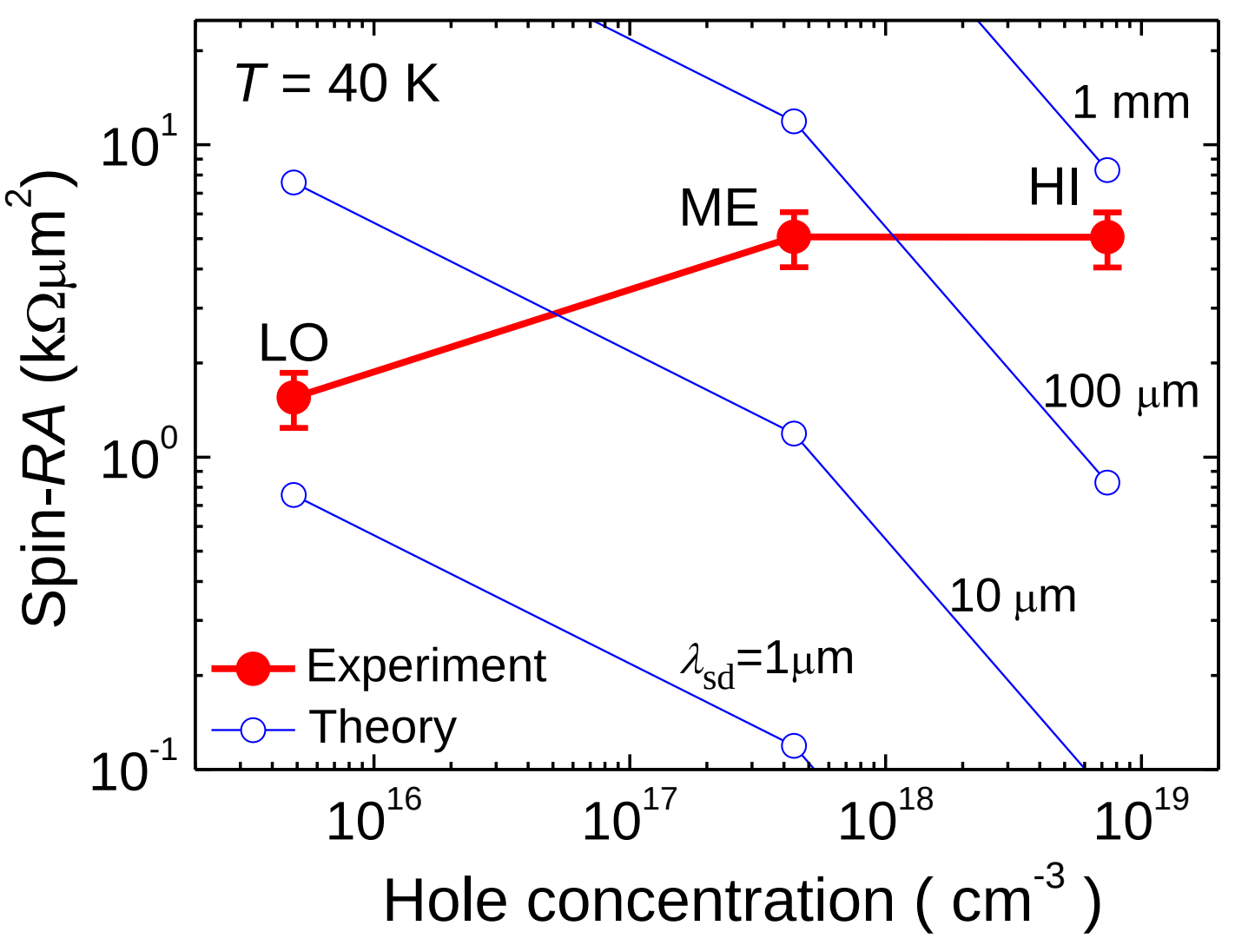

Fig.4 Iba *et al*.